\documentclass[letterpaper]{article} % DO NOT CHANGE THIS
\usepackage{aaai25}  % DO NOT CHANGE THIS
\usepackage{times}  % DO NOT CHANGE THIS
\usepackage{helvet}  % DO NOT CHANGE THIS
\usepackage{courier}  % DO NOT CHANGE THIS
\usepackage[hyphens]{url}  % DO NOT CHANGE THIS
\usepackage{graphicx} % DO NOT CHANGE THIS
\usepackage{natbib}  % DO NOT CHANGE THIS AND DO NOT ADD ANY OPTIONS TO IT
\usepackage{caption} % DO NOT CHANGE THIS AND DO NOT ADD ANY OPTIONS TO IT
\usepackage{algorithm}
\usepackage{algorithmic}

\usepackage{newfloat}
\usepackage{listings}
\DeclareCaptionStyle{ruled}{labelfont=normalfont,labelsep=colon,strut=off} % DO NOT CHANGE THIS
\floatstyle{ruled}
\newfloat{listing}{tb}{lst}{}
\floatname{listing}{Listing}
\usepackage{amsmath,amsfonts, amssymb, subcaption, booktabs}

\title{A Learning Framework For Cooperative Collision Avoidance of UAV Swarms Leveraging Domain Knowledge}
\author {
    Shuangyao Huang \textsuperscript{\rm 1}, 
    Haibo Zhang \textsuperscript{\rm 2}, 
    Zhiyi Huang \textsuperscript{\rm 2} 
}
\affiliations {
    \textsuperscript{\rm 1} School of Internet of Things, Xi'an Jiaotong-Liverpool University, Suzhou 215400, China \\
    \textsuperscript{\rm 2} School of Computing, University of Otago, Dunedin 9016, New Zealand \\
    shuangyao.huang@xjtlu.edu.cn, haibo.zhang@otago.ac.nz, zhiyi.huang@otago.ac.nz
}
\usepackage{bibentry}
\begin{document}

\maketitle

\begin{abstract}
	This paper presents a multi-agent reinforcement learning (MARL) framework for cooperative collision avoidance of UAV swarms leveraging domain knowledge-driven reward. The reward is derived from knowledge in the domain of image processing, approximating contours on a two-dimensional field. By modeling obstacles as maxima on the field, collisions are inherently avoided as contours never go through peaks or intersect. Additionally, counters are smooth and energy-efficient. Our framework enables training with large swarm sizes as the agent interaction is minimized and the need for complex credit assignment schemes or observation sharing mechanisms in state-of-the-art MARL approaches are eliminated. Moreover, UAVs obtain the ability to adapt to complex environments where contours may be non-viable or non-existent through intensive training. Extensive experiments are conducted to evaluate the performances of our framework against state-of-the-art MARL algorithms. 
\end{abstract}

 \begin{links}
%	\link{Code}{https://github.com/sunningshore/reMARL.git}
	%  \link{Datasets}{https://aaai.org/example/datasets}
	%  \link{Extended version}{https://aaai.org/example/extended-version}
	Under review at AAAI 2026
\end{links}

\section{Introduction} \label{IT}

We investigate the cooperative collision avoidance problem in multi-rotor unmanned aerial vehicle (UAV) swarms. A UAV swarm refers to a fleet of UAVs working collaboratively to accomplish one mission. 
%Compared to single UAVs, swarms exhibit significantly enhanced performance across various applications such as delivery, communication relay, search and rescue, etc. 
%For instance, they improve delivery efficiency through distributed load-carrying, expand the coverage area for post-disaster wireless communication via distributed relaying, enhance the precision and accuracy of search and rescue missions through perceptual diversity and sensor fusion technology, and extend flight time via collaborative aerial charging. 
%However, the challenge of effective collision avoidance prevents the widespread of these applications. 
The primary difficulty lies in ensuring seamless cooperation among swarm members. The key objectives for collision avoidance in UAV swarms are safety and energy efficiency, as the flight time of off-the-shelf UAVs is constrained by their limited onboard power supply. 

Conventional approaches based on velocity obstacle (VO) \cite{reciprocalvo}, artificial potential field (APF) \cite{b11}, and meta-heuristic optimization \cite{hyb6} 
%\cite{hyb5, hyb6} 
suffer from low energy efficiency due to their lack of long-term planning. 
%the short-sightedness and lack of cooperation. These approaches adopt a receding time horizon planning approach, where UAVs plan trajectories considering conditions within a limited time window which is shifted along time horizon. As a result, the trajectories are only viable within the current time window and will be frequently adjusted across time windows, which is energy consuming. 
Moreover, these approaches are difficult to achieve cooperation among UAVs as they regard each other as part of the environment. 
%As a result, the UAVs keep avoiding each other resulting in zigzag trajectories which are energy inefficient. 
Another class of hybrid methods \cite{huang2021, huang2023} integrate APF with particle swarm optimization (PSO), a meta-heuristic optimization method, to address the above limitations. 
%find trajectories that approximate contours on the potential field can achieve energy efficiency and long-term optimality in collision avoidance. By modeling obstacles as maxima on the field, collisions are avoided as contours never go through peaks or intersect. Moreover, counters are smooth and energy-efficient. 
However, meta-heuristic optimization-based approaches are limited by long reaction time, and therefore, are not suitable for real-time UAV applications in complex environments. 
%Single agent reinforcement learning (RL) models such as DDPG \cite{ddpg} and DQN \cite{dqn} can make real-time decisions but struggle to learn cooperative models in multi-agent settings, due to the challenges of `lazy agents' and non-stationary environment in multi-agent settings \cite{lazy-agent, non-stationarity}. 

Recent advances in multi-agent reinforcement learning (MARL) models address the limitations of conventional approaches in UAV swarm collision avoidance, through the techniques of observation sharing, credit assignment, and reward engineering. 
%problems of short-sightedness, long reaction time, and ineffective cooperation. 
%However, existing models have limitations in UAV operations. 
Early research 
%such as Multi-Agent Deep Deterministic Policy Gradient (MADDPG) \cite{MADDPG} and Independent $Q$ Learning (IQL) \cite{iql} 
rely on sharing observations among swarm members to achieve cooperation. 
%In these approaches, the input to each agent's network contains the observations of all agents, either simultaneously or alternately. 
This method shows degraded performances when the number of agents increases. Moreover, the sharing frequency becomes an additional hyperparameter that affects the performances, as observed in IQL \cite{iql}. Alternative approaches employ sophisticated credit assignment mechanisms to facilitate cooperation among agents under a team reward. However, explicit credit assignment-based approaches are usually designed for discrete action spaces, making it unsuitable for UAV operations that require continuous control outputs, such as COMA \cite{COMA}. Implicit credit assignment-based approaches, on the other hand, lead to unbounded divergence and unexpected actions \cite{divergence}, such as VDN \cite{VDN} and QMIX \cite{Qmix}. Moreover, these sophisticated credit assignment mechanisms become ineffective when the number of agents increases, as observed in CoDe \cite{huang2024}. Reward engineering enables training with large swarm sizes by shifting the burden with facilitating cooperation from network structure to reward design, either through imitation learning or exploiting domain knowledge. 
Imitation learning minimizes the discrepancy between agent behaviors and expert demonstrations \cite{mail2} 
%\cite{mail1, mail2}
, making its effectiveness inherently dependent on the quality of available demonstrations. Moreover, imitation learning typically cannot ensure robustness against deviations by strategic agents. When agents operate under alternative utility functions, the resulting state distributions may diverge from those observed in the demonstrations \cite{mail4}. 
Domain knowledge reduces reliance on high-quality demonstrations and minimizes risks from strategic deviations. However, few studies have explored reward engineering using domain knowledge, whether in UAV swarm missions or interdisciplinary fields. We argue that insights from interdisciplinary domains can be adapted to facilitate training with large UAV swarms in cooperative collision avoidance. 
%Although most current research in MARL are dedicated to investigating observation sharing, credit assignment, and imitation learning paradigms, few research has investigated reward engineering with domain knowledge for training large UAV swarms in cooperative collision avoidance that eliminates the need for high-quality demonstrations and mitigates risks associated with strategic deviations. 

%An example of policy-based models is Counterfactual Multi-Agent Policy Gradients (COMA) \cite{COMA}, which explicitly estimates the individual contribution of each agent's actions through counterfactual reasoning. However, COMA is specifically designed for discrete action spaces, making it unsuitable for UAV operations that require continuous control outputs. Value-based models such as VDN \cite{VDN} and QMIX \cite{Qmix} employ implicit credit assignment schemes by decomposing the swarm's action value into individual utility functions. This value decomposition leads to unbounded divergence and unexpected actions \cite{divergence}. Moreover, these approaches become ineffective when the number of agents increases.  Although most current research in MARL are dedicated to investigating observation sharing and credit assignment schemes, few research has investigated reward engineering. This paper proposes a novel MARL approach for cooperative collision avoidance of UAV swarms based on reward engineering. 

This paper proposes \textit{reMARL}, a MARL framework for cooperative collision avoidance of UAV swarms leveraging domain knowledge-driven reward. The reward function is designed by leveraging principles from image processing, where UAV trajectories are approximated as contours on a two-dimensional field with obstacles represented as peaks. This integration inherently ensures safety and energy efficiency of UAVs, as contours never go through field peaks or intersect, and the trajectories are smooth and continuous curves. 
%The framework features an innovative integration of a meta-heuristic optimization approach with MARL, where the meta-heuristic component provides reward signals defined by domain knowledge within the MARL framework. 
%aiming to guide the UAVs to learn safe, energy-efficient, and cooperative actions in collision avoidance.  
%This paper proposes \textit{reMARL} - an innovative integration of a contours-based meta-heuristic optimization approach with MARL through reward engineering. The meta-heuristic component provides domain knowledge-driven reward signals within the MARL framework, aiming to guide the UAVs to learn safe, energy-efficient, and cooperative actions in collision avoidance. 
%Specifically, the reward signals in MARL are designed using the cost function derived from meta-heuristic optimization. 
Compared with state-of-the-art MARL approaches, our framework enables training with large swarm sizes by minimizing the agent interaction in two ways:  
\begin{enumerate}
	\item Observation sharing is only used in constructing rewards, decoupling it from network optimization and avoiding the issues that arise from directly applying observation sharing to networks. 
	\item The need for complex credit assignment schemes is eliminated as UAVs can learn cooperative behaviors solely by maximizing their individual rewards. 
\end{enumerate}
Additionally, maximizing these reward ensures that the UAVs' trajectories approximate contours on a two-dimensional field, thereby stabilizing and bounding their actions. On the other hand, the UAVs obtain the ability to adapt to complex environments where contours may be non-viable or non-existent through intensive learning. 
%Lastly, the reaction time is substantially reduced, as actions are determined through a simple forward pass of neural networks. 

\section{Related Work} \label{RW}

Works \cite{vo_oda1, vo_oda2} explores the VO method in UAV swarm collision avoidance. These works avoid collisions by adjusting the UAV's velocity. 
%outside a predefined pool of velocities that would result in collisions with other UAVs or obstacles. While straightforward, these methods operate under the assumption that obstacles can be approximated as circles, with safety distances serving as their radii. Consequently, these methods struggle in complex environments where obstacles exhibit irregular shapes and cannot be accurately modeled as circles. 
While straightforward, this method induces frequent velocity change and is energy-inefficient.  
Works \cite{apf_oda1, b11} addresses the collision avoidance of UAVs based on APF. These works offer a more versatile solution by modeling complex environments as two-dimensional differentiable potential fields, irrespective of the number or shapes of obstacles. Once the environment is modeled, UAVs are guided to avoid collisions by following the gradients of the potential field. However, a significant drawback of these works is the presence of local optima on the potential field where gradients vanish. These local optima can trap UAVs, leading to energy-inefficient, oscillatory trajectories. 
Works \cite{geneticapplication1, pathfinding3} adopt meta-heuristics methods which address energy efficiency in collision avoidance by minimizing cost functions that account for energy consumption. 
%\cite{hyb5}, \cite{geneticapplication1}, and \cite{pathfinding3} explore swarm intelligence, genetic algorithms, and graph path-finding algorithms, respectively. These works address energy efficiency in collision avoidance by minimizing cost functions that account for energy consumption. 
However, they are hindered by the curse of dimensionality. As the size of the swarm increases, the dimensionality of the search space expands and the collaboration among swarm members becomes complicated. Works \cite{huang2021, huang2023} exploit image processing domain knowledge in UAV swarm collision avoidance by letting UAVs fly along contours on a shared field. The optimal trajectories are found using PSO, a meta-heuristic search method, and therefore, is limited by time complexity for UAV operations. 

Recent works leverage MARL models to train cooperative collision avoidance strategies for UAVs. Works \cite{iql_oda1} and \cite{MADDPG_oda1} explore IQL and MADDPG \cite{MADDPG} models, respectively. These works rely on observation sharing among agents, and consequently facing exponentially increasing observation space dimensionality as swarm sizes increase. 
%These works rely on observation sharing among agents, and hereby often produce sub-optimal policies due to the lack of effective credit assignment mechanisms, which are crucial for accurately attributing rewards to individual agents in cooperative missions. Moreover, these works necessitate UAV-to-UAV communication in execution. 
In contrast, other works \cite{vdn_oda1, qmix_oda1, coma_oda1} investigate credit assignment models such as VDN, QMIX, COMA, and outperforms prior works based on observation sharing. Explicit credit assignment models such as COMA introduce a centralized value network that takes the joint observations and actions of all agents and other global information as input, and derives a baseline for each agent that measures its contribution from the perspective of other agents. However, this centralized value network induces scalability issues as the input dimension grows exponentially as the swarm size increases. Implicit credit assignment models such as VDN and QMIX, on the other hand, achieve a balance between scalability and learning stability. These models are based on assumptions such as Individual-Global-Max (IGM) which states that maximizing the global action value is equivalent to maximizing individual local action values. Although IGM simplifies the training with large swarm sizes as actions can be selected locally rather than maximizing the global value, this simplicity can result in unbounded divergence and unpredictable agent behaviors \cite{divergence}. The work in \cite{mail2} employs reward engineering through a two-stage approach, where a policy network is first trained via supervised learning using expert demonstrations generated by VO, and is then enhanced through reinforcement learning. However, the resulting models are optimized to imitate VO, leading to energy inefficiency. Additionally, this two-stage approach suffers from strategic deviations in agent behavior.

\section{Background} \label{BG}

\subsection{Domain Knowledge For Cooperative Collision Avoidance} 

Leveraging image processing domain knowledge in cooperative collision avoidance for UAV swarms is first proposed in work \cite{huang2021}. The key idea is to model the environment as a two-dimensional potential field and design a cost function that optimizes a curve to approximate contours on the field. This cost function is designed leveraging the advantages of the active contour model \cite{snakes}, a common approach for contour extraction in image processing. 
%For energy considerations, these methods confine the UAVs' trajectories during collision avoidance in two-dimensional space as altitude changes consume much more energy than level flight \cite{huang2021}. 

The potential field is constructed as the superposition of multiple repulsive fields, where each obstacle is represented as an individual repulsive field. Additionally, the UAV swarm is treated as a single entity and is also modeled as a repulsive field. The intensity at position $q$ on the repulsive field for an obstacle is defined as follows. 
\begin{equation} \label{eq:obstaclefield} 
	\begin{aligned} 
		\varPhi_o(q)=\left\{\begin{array}{c c}  
			\dfrac{\max\{v_o, v_s\}}{d^2_{safe}} & |{q, p_o}|\leq d_{safe} \\ 
			\dfrac{\max\{v_o, v_s\}}{|{q, p_o}|^2} & d_{safe} < |{q, p_o}| \leq R_o \\ 
			0 & |{q, p_o}|> R_o 
		\end{array} 
		\right. , 
	\end{aligned} 
\end{equation} 
where $p_o$ denotes the position of the obstacle, and $R_o$ represents the influential range of the repulsive field. $v_o$ and $v_s$ are the obstacle's and swarm's velocities, respectively. The $\max$ operator ensures that the intensities of the swarm field and the obstacle's field remain comparable when $v_o<v_s$, particularly in cases where $v_o=0$ for static obstacles. The parameter $d_{safe}$ defines the minimum safe distance that a UAV must maintain from obstacles to ensure safety. The region within $d_{safe}$ around the obstacle exhibits the maximum field intensity, effectively preventing UAVs from approaching closer than the specified safety threshold. 

On the other hand, the intensity at position $q$ on the repulsive field of the swarm is defined as follows. 
\begin{equation} \label{eq:swarmfield} 
	\begin{aligned} 
		\varPhi_s(q)=\left\{\begin{array}{c c}  
			\dfrac{v_s}{|{q, p^*}|^2} & |{q, p^*}| \leq R_s \\ 
			0 & |{q, p^*}| > R_s
		\end{array} 
		\right. , 
	\end{aligned} 
\end{equation} 
where $R_s$ denotes the influential range of the repulsive field. The term $p^*$ represents the virtual center of the swarm, which is a conceptual point positioned ahead of the swarm. Rather than modeling each UAV individually, the swarm is collectively represented as a single repulsive field. This approach prevents UAVs from being positioned at peaks on their individual repulsive fields, where contours are non-existent. The overall potential field is defined as 
\begin{equation} \label{eq:environmentfield} 
	\begin{aligned} 
		\varPhi(q)=&\varPhi_s(q)+ \sum\varPhi_o(q). 
	\end{aligned} 
\end{equation} 
In this way, obstacles and the swarm virtual center are placed at field peaks. 

The cost function is built upon the potential field and is defined as follows. 
\begin{equation} \label{eq:fitness} 
	\begin{aligned} 
		f(S(\rho))=\int \dfrac{1}{2}|S''(\rho)|^2 - \dfrac{1}{2}|\triangledown\varPhi(S(\rho))|^2 d\rho, 
	\end{aligned} 
\end{equation} 
where $S(\rho)$ represents the trajectory to be optimized and $\rho$ is an arc length parameter. The cost function comprises two components: an integral part and a differential part. The integral part optimizes a trajectory's internal properties by maximizing its continuity and smoothness. The differential part acts as an edge detector, ensuring safety by aligning the trajectory with the edges of the potential field. Eq. \eqref{eq:fitness} resembles the active contour model, with the only difference being that a term for minimizing trajectory length is missing in our model as the trajectory length is fixed to the UAV's velocity. Thereby, trajectories minimizing Eq. \eqref{eq:fitness} approximate contours on the potential field $\varPhi$. In this way, safety of UAVs is ensured as contours never go through field peaks or intersect, and energy efficiency is optimized as contours are smooth and continuous trajectories. 

\subsection{Markov Games} 

In MARL, $n$ agents are trained by interacting with each other in a dynamic environment through a sequence of actions, observations, and rewards. Each sequence is assumed to terminate within a finite number of time steps. This framework induces a finite Markov game. A Markov game is defined by a tuple $(\mathcal{S}, \mathcal{A}, \mathcal{O}, \{r_i\}_{i=1}^{n}, \mathcal{P})$, representing the state space, action space, observation space, reward signals, and state transition probability of the environment. Specifically, the state transitions and reward signals are determined by the joint actions of agents $\mathcal{A} = \mathcal{A}_1 \times \cdots \times\mathcal{A}_n$. At each step, each agent makes an observation $o_i\in \mathcal{O}$ on the environment, and selects an action $a_i\in \mathcal{A}_i$ based on its observation. Additionally, the environment also has a true state $s \in \mathcal{S}$. 
When all the agents execute their actions simultaneously, the environment transits from state $s$ to $s'$ with probability $P(s'|s, \boldsymbol{a}): \mathcal{S}\times\mathcal{A}\times\mathcal{S}\rightarrow[0,1]$, and returns a set of numerical rewards $\{r_i\}_{i=1}^{n}: \mathcal{S}\times\mathcal{A}\rightarrow\mathbb{R}^n$, where $\boldsymbol{a}=\{a_i\}_{i=1}^{n}$ denotes the joint action of all agents. 

\subsection{MARL} 
Value-based models such as VDN and QMIX follow DQN architecture, where a value network $\phi$ is trained to approximate the action value $Q(s, a)$. Cooperation is achieved through decomposing the joint action value for the entire swarm to individual utility values of agents. Each agent maintains a value network $\bar{Q}_i(s, a_i)$. These individual values are composed into the joint action value $Q(s, \boldsymbol{a})$, which is updated by minimizing a temporal difference loss on the transition $\{s, a, r, s'\}$ sampled from an experience buffer $\mathcal{D}$: 
\begin{equation} \label{tdloss} 
	\begin{gathered} 
		\mathcal{L} = \mathbb{E}_{\mathcal{D}}\left[(y - Q(s, \boldsymbol{a}))^2\right],  
	\end{gathered} 
\end{equation} 
where $y = r + \gamma \cdot Q(s', \boldsymbol{a}')$ and $a'=\mathop{\arg\max}\limits_{\Tilde{a}\in\mathcal{A}} Q(s', \Tilde{a})$. 
%For partial observability, $s$ is replaced with $o$. 
%and $\bar{\omega}$ represents the parameters of a target value network \cite{double-q-learning}, which are periodically copied from $\omega$. The target value network shares the same architecture as the value network. The purpose of the target critic is to stabilize and bound the estimation of the $Q$ value. 

Policy-based models such as COMA and CoDe follow actor-critic architecture \cite{actor-critic}, where a value network $\psi_{\omega}$ is trained to approximate the action value $Q$ for the entire swarm and a policy network $\pi_{\theta}$ is trained to render action for each agent. The value network is updated by minimizing the temporal difference loss Eq. \ref{tdloss}. On the other hand, the policy network is updated by policy gradients: 
\begin{equation} \label{policygradient} 
	\begin{gathered} 
		g=\mathbb{E}_{\pi_{\theta}} \left[\nabla_{\theta}\log\pi_{\theta}\cdot A_{\omega}(\theta)\right], 
	\end{gathered} 
\end{equation}
where $A_{\omega}$ is an agent-specific advantage function that measures an agent's contribution in acquiring the value $Q$ based on the concept of difference rewards \cite{diffreward}. 
This advantage function is derived from the value network and hence parameterized by $\omega$ and $\theta$. 
%The advantage function is defined by subtracting a counterfactual baseline from the swarm's action value in policy-based approaches, and equals individual action value in value-based approaches. 

%For works based on the network architecture of DQN \cite{dqn}, such as VDN and QMIX, the values and policies are approximated using one network $Q_{\omega}(s, a)$, which is updated by minimizing the TD loss: 
%\begin{equation} \label{criticloss1} 
%	\begin{gathered} 
	%		\mathcal{L}_{TD} = \mathbb{E}_{\{s, a, r, s'\}\sim D}\left[
	%		\left(y - \sum_{i}\alpha_i \cdot Q_{\omega}(s, a|i)\right)^2\right],  
	%	\end{gathered} 
%\end{equation} 
%where $y = r + \gamma \cdot \sum_{i} \alpha_i \cdot \max_{a'} Q{\bar{\omega}}(s', a'|i)$, $Q(\cdot|i)$ represents the individual action value decomposed from the global value, $\alpha_i$ denotes the weight in decomposition, and the transitions $\{s, a, r, s'\}$ are sampled from the experience buffer $D$. 

%Eventually, execution is performed online and it only requires the policy networks to generate actions. 

%\begin{figure*}[htbp] 
%	\centering
%	\begin{subfigure}[b]{.48\linewidth}
	%		\centering
	%		\includegraphics[width=1.0\linewidth]{figures/frame1.pdf} 
	%		\caption{} 
	%		\label{1a} 
	%	\end{subfigure} 
%	\begin{subfigure}[b]{.48\linewidth}
	%		\centering
	%		\includegraphics[width=1.0\linewidth]{figures/frame2.pdf} 
	%		\caption{} 
	%		\label{1b} 
	%	\end{subfigure} 
%	\caption{(a) conventional MARL framework based on credit assignment schemes and (b) the proposed MARL framework with contours-based reward engineering. } 
%	\label{frameworks} 
%\end{figure*} 

\begin{figure*}[htbp] 
	\centering 
	\includegraphics[width=.8\linewidth]{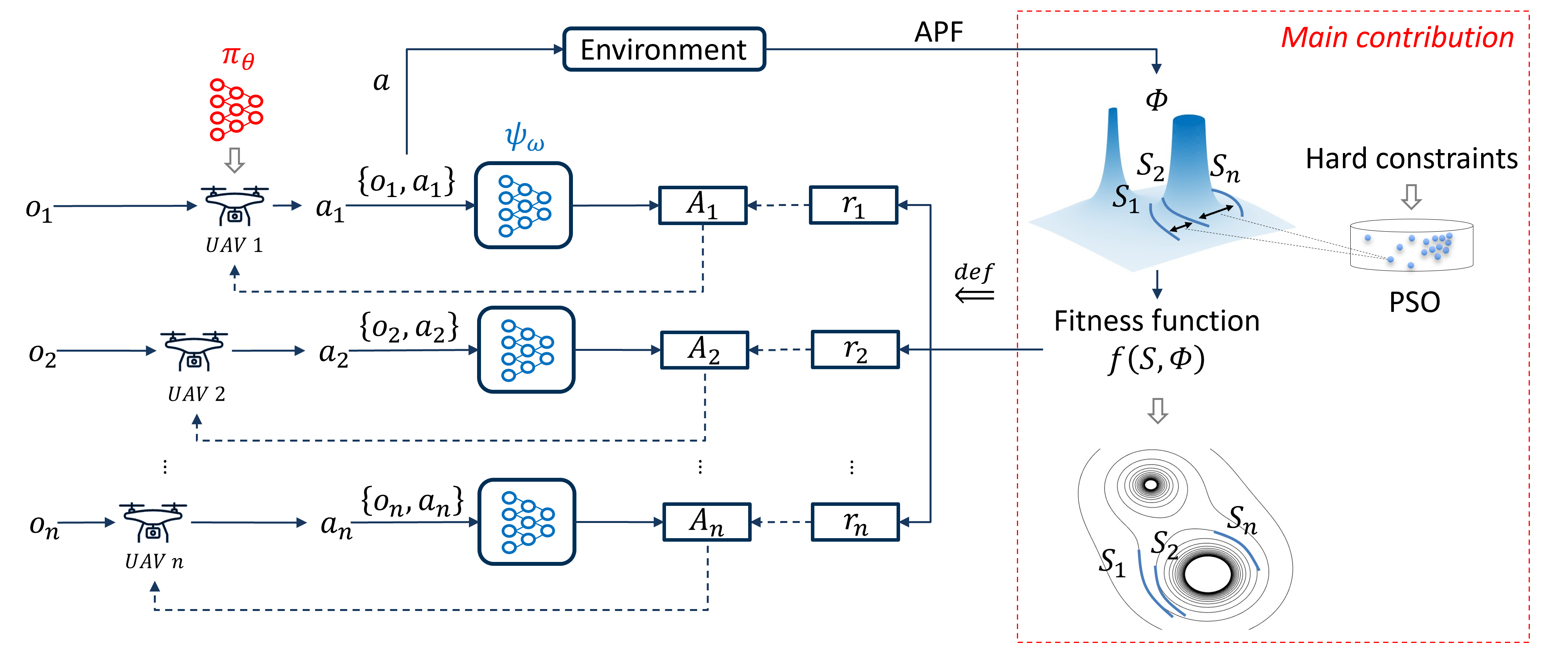} 
	\caption{The proposed MARL framework with domain knowledge-driven reward. } 
	\label{frameworks} 
\end{figure*} 

\section{Application Scenario} 

%To align with the settings in contour-based meta-heuristic optimization, we restrict the movement of UAVs during collision avoidance to a two-dimensional space. 
For energy considerations, these methods confine the UAVs' trajectories during collision avoidance in two-dimensional space as altitude changes consume much more energy than level flight \cite{huang2021}. 
Consequently, the observation of a UAV includes only the  $x-$ and $y-$ components of its kinematic properties. For simplicity, the action of a UAV is defined as the change in its velocity direction. To ensure smooth trajectories, this change is constrained to the range $[-{\pi}/{4}, {\pi}/{4}]$. 
In addition, we assume the magnitudes of the UAVs' velocities remain constant during collision avoidance for energy efficiency. This assumption is based on the rationale that UAV collision avoidance typically occurs within a short timeframe, and changing velocity magnitudes during this short interval would require significant acceleration, leading to high energy consumption. 
Lastly, due to the constraints imposed by the limited aperture angle and sensing range of LiDAR sensors, UAVs can only observe a part of the environment. This limitation is referred to as partial observability. 

\section{A MARL Framework With Domain Knowledge-Driven Reward} 

\subsection{Framework} 

%Building on contours-based meta-heuristic optimization and MARL, we introduce an innovative reward design that guides UAVs to find contour-like trajectories which are cooperative, safe, and energy-efficient. 
%The proposed framework in comparison with conventional MARL framework based on credit assignment schemes is illustrated in Fig. \ref{frameworks}. In Fig. \ref{frameworks}, we refer to the global action value of the swarm as $Q$, and individual action values or advantage functions as $A$. 
The proposed framework is illustrated in Fig. \ref{frameworks}, where we refer to the global action value of the swarm as $Q$, and individual action values or advantage functions as $A$. 
%The conventional MARL framework relies on credit assignment schemes to devise an advantage function for each agent from the global action value which is updated by a team reward. On the contrary, 
The proposed framework first maps the environment to a potential field, upon which, an individual reward is defined for each agent using Eq. \eqref{eq:fitness} that guides the UAVs to fly on contours. PSO is used to adjust each UAV's contour level such that the distance between contours are larger than a threshold as a hard constraint. 
This reward is directly used to update the individual action value for each agent. Therefore, each agent can have its own policy and value network, and only needs to consider its own individual reward. As these rewards are derived from a common potential field using the active contour model, maximizing these rewards automatically results in cooperative, safe, and energy-efficient trajectories. The component in a red dashed square defines the reward signals and is our main contribution. 

\subsection{Innovative Reward Design} 

%We keep in mind three principles when designing the reward for cooperative collision avoidance: 
%\begin{itemize}
%	\item Formation principle: UAVs must maintain proximity to one another while avoiding collisions for uninterrupted mission and communication. 
%	\item Efficiency principle: UAVs should conserve energy by minimizing unnecessary velocity changes during avoidance. 
%	\item Safety principle: UAVs must ensure safety by avoiding collisions with obstacles and each other. 
%\end{itemize} 
Concerns of cooperation, formation, and safety of the UAVs must be promoted when designing the reward. Moreover, the reward should prioritize safety over the other two concerns. Overall, the reward consists of two parts: 
%a \textit{Contour} part which is our main contribution and achieves cooperation in collision avoidance by planning trajectories on contours and a \textit{Swarming} part which is regarded as a general term for swarming and ensures safety and formation. 
\begin{itemize}
	\item A \textit{Contour} part which is our main contribution and achieves cooperation in collision avoidance by planning trajectories on contours, and 
	\item A \textit{Swarming} part which is regarded as a general term for swarming and ensures safety and formation. 
\end{itemize}
The reward is defined as follows. 
\begin{equation} \label{eq:reward} 
	\begin{aligned} 
		r = \underbrace{-f(S, \varPhi(q))}_\text{\textit{Contour}} + \underbrace{r_{form} \cdot r_{collide}}_\text{\textit{Swarming}}, 
	\end{aligned} 
\end{equation} 
%where $r_{form}$ and $r_{collide}$ enforce the formation and safety principles, respectively.
where, 
%$r_{form}=0$ if the UAV deviates from its pre-planned trajectory beyond a threshold distance $d_{form}$, and $r_{form}=1$ otherwise. 
$r_{form}=\dfrac{\langle \boldsymbol{v}, \bar{\boldsymbol{v}} \rangle}{||\boldsymbol{v}||\cdot||\bar{\boldsymbol{v}}||}$, with $\boldsymbol{v}$ and $\bar{\boldsymbol{v}}$ representing the UAV's real-time velocity avoiding collisions and pre-planned velocities before collision avoidance, respectively. In this way, $r_{form}$ promotes formation of UAVs within a swarm by minimizing unnecessary velocity changes during avoidance. On the other hand, $r_{collide}$ acts as a safety reward, with $r_{collide}=0$ if a collision occurs (when $d_{U2O}$ falls below a threshold) and $r_{collide}=1$ otherwise. Safety is prioritized as the second term of Eq. \ref{eq:reward} has positive values only when the UAVs avoid collisions. Otherwise, it is zero. 

\begin{figure}[bt] 
	\centering
	\begin{subfigure}[b]{.48\linewidth}
		\centering
		\includegraphics[width=1.0\linewidth]{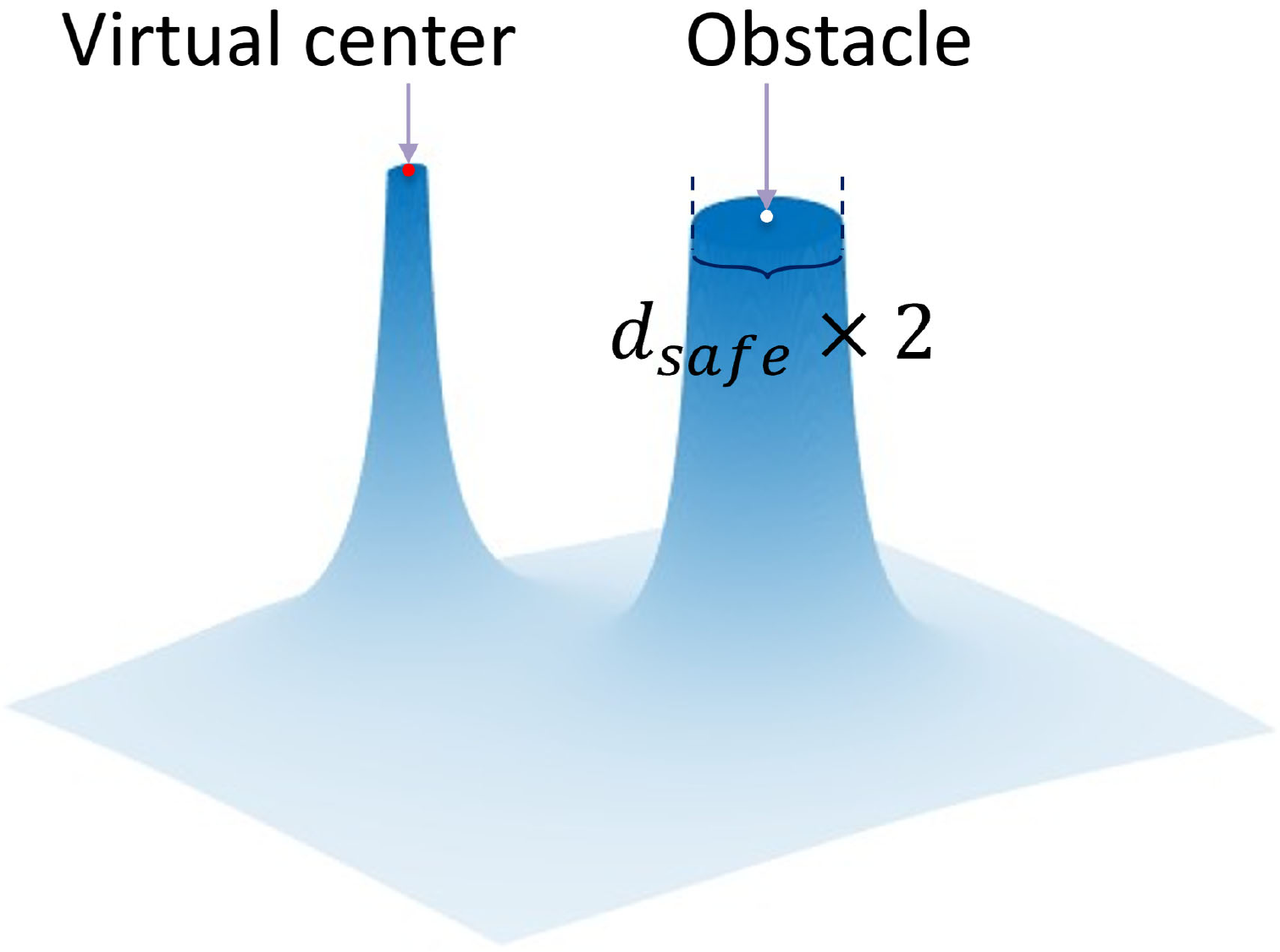} 
		\caption{} 
		\label{fieldsa} 
	\end{subfigure} 
	\begin{subfigure}[b]{.48\linewidth}
		\centering
		\includegraphics[width=1.0\linewidth]{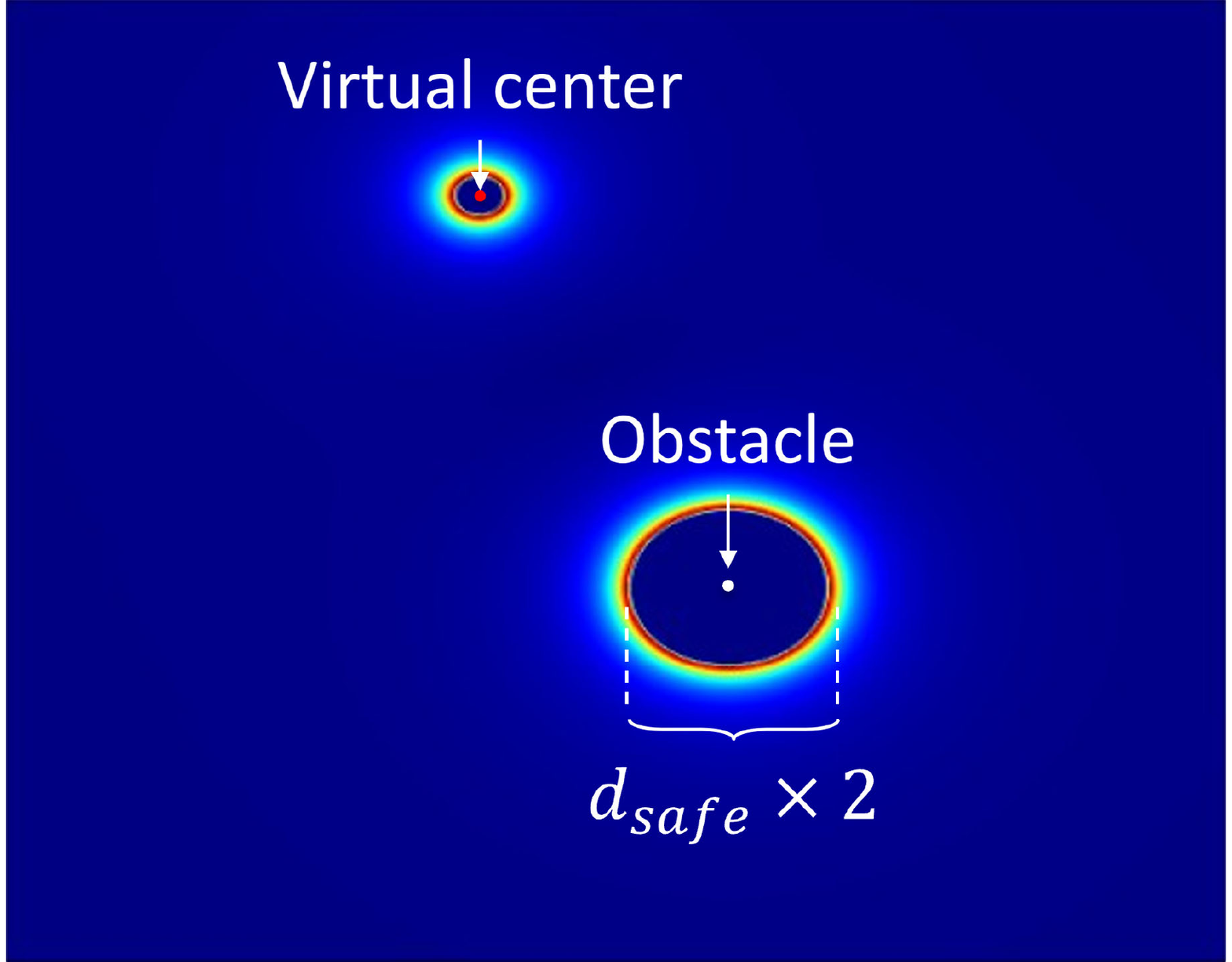} 
		\caption{} 
		\label{fieldsb} 
	\end{subfigure} 
	\caption{Potential field $\varPhi$ mapped from the environment and its gradient amplitudes. } 
	\label{fields} 
\end{figure}

The cost function $f(S, \varPhi)$ for a random curve $S$ on the potential field is defined using Eq. \eqref{eq:fitness}. 
%and requires observation sharing among agents. This is the only place observation sharing is required. 
We first map the environment to a potential field $\varPhi$ using Eq. \ref{eq:environmentfield} as shown in Fig. \ref{fieldsa}. The edges at $d_{safe}$ away from the obstacle have the maximum gradient amplitudes around the obstacle as shown in Fig. \ref{fieldsb}. Therefore, the cost function $f(S, \varPhi)$ achieves minimum values when the curve $S$ locates on the edges, encouraging the UAVs to keep a safe distance from the obstacle. 

In addition to separating UAVs and obstacles, it is necessary to ensure hard constraints. For instance, the minimum distance between any UAV and obstacle ($d_{U2O}$) and between any two UAVs ($d_{U2U}$) should be larger than a threshold $\bar{d}_{U2O}$ and $\bar{d}_{U2O}$, respectively, at any time. The hard constraint on $d_{U2O}$ is ensured by $r_{collide}$ in the \textit{Swarming} part. On the other hand, $d_{U2U}$ is ensured by adjusting the position of UAVs in their vicinity, such that their contours are at least $d_{U2U}$ away from each other. The position of UAVs determine their contours on the potential field, and must be adjusted together. This collective adjustment is achieved with PSO simply by minimizing the following cost function. 
\begin{equation} \label{eq:psocost} 
	\begin{aligned} 
		f_{pso} = f_{thres} + f_{shift}, 
	\end{aligned} 
\end{equation} 
where 
\[	
f_{thres} = \begin{cases} d_{U2O}, & d_{U2O} \geq \bar{d}_{U2O},\\
	\text{Inf},  &  \text{otherwise},
\end{cases}
\]
and $f_{shift}=\max_{i \in [1, N]}||q_i, q'_i||$, where $q_i$ and $q'_i$ are the position of UAV $i$ before and after adjustment, respectively. $N$ is the number of UAVs in the swarm. Minimizing $f_{thres}$ ensures that the minimum distance between any two UAVs remains either slightly greater than or equal to the threshold $\bar{d}_{U2O}$, while minimizing $f_{shift}$ reduces the maneuver effort of UAVs for efficiency. 

The optimal solution that minimizes Eq. \ref{eq:psocost} is found following standard PSO optimization. A swarm of particles $\boldsymbol{\xi}^{0}=\{\xi_i^{0}|i=1,2,\cdots\}$ is first initialized, each representing the position of $N$ UAVs $\xi_i^0=[(p_{s|x}, p_{s|y})_1, \cdots, (p_{s|x}, p_{s|y})_N]$. The position of particle $i$ is updated in an iterative manner following Eq. \ref{eq:pso}. 

\begin{equation} \label{eq:pso} 
	\begin{aligned} 
		&\boldsymbol{v}_i{'} \leftarrow \mu\boldsymbol{v}_i+c_1 ({p}_i-{\xi}_i)+c_2 ({p}_g-{\xi}_i), \\ 
		&{\xi}_i{'}\leftarrow{\xi}_i+\boldsymbol{v}_i, 
	\end{aligned} 
\end{equation} 
where $\boldsymbol{v}_i$ is the update vector for particle $i$, ${p}_i$ and ${p}_g$ are the personal best position for particle $i$ and the global best position for all particles, respectively. Coefficient $\mu$ is an inertia weight uniformly distributed in $[0, 1]$, and $c_1$, $c_2$ are acceleration coefficients for cognitive and social components. The time complexity of this heuristic search does not impact the algorithm's real-time execution as it's only needed in training. 

%The optimal solution that minimizes Eq. \ref{eq:psocost} is found following standard PSO optimization illustrated in Algorithm \ref{algo1}. This heuristic search is only needed in training. Therefore, its time complexity does not impact the algorithm's real-time execution. 
%\begin{algorithm}
%	\caption{PSO for UAV Position Adjustment} \label{algo1}
%	\begin{algorithmic}[1]
	%		\STATE \textbf{Initialization}: Initialize a swarm of particles $\boldsymbol{\xi}^{0}=\{\xi_i^{0}|i=1,2,\cdots\}$. Each particle represents the position of $N$ UAVs $\xi_i^0=[(p_{s|x}, p_{s|y})_1, \cdots, (p_{s|x}, p_{s|y})_N]$. Randomly initialize the update vector $\boldsymbol{v}_i^{0}$ for each particle, an inertia weight $\mu$ uniformly distributed in $[0, 1]$, and acceleration coefficients $c_1$, $c_2$ for cognitive and social components. 
	%		\FOR{$ite$ in 1,2, ...}
	%			\STATE Calculate the cost for each particle using Eq. \ref{eq:psocost}. 
	%			\STATE Find the $p$-best (${p}_i^{ite}$) and $g$-best (${p}_g^{ite}$) particles. 
	%			\STATE Update the particles by 
	%			\FOR{$i$ in 0,1,2, ...} 
	%				\STATE $\boldsymbol{v}_i^{ite} \leftarrow \mu\boldsymbol{v}_i^{ite-1}+c_1 ({p}_i^{ite}-{\xi}_i^{ite-1})+c_2 ({p}_g^{ite}-{\xi}_i^{ite-1}). $ 
	%				\STATE ${\xi}_i^{ite}\leftarrow{\xi}_i^{ite-1}+\boldsymbol{v}_i^{ite}. $
	%			\ENDFOR
	%		\ENDFOR
	%	\end{algorithmic}
%\end{algorithm}

\begin{figure}[bt] 
	\centering
	\begin{subfigure}[b]{.48\linewidth}
		\centering
		\includegraphics[width=1.0\linewidth]{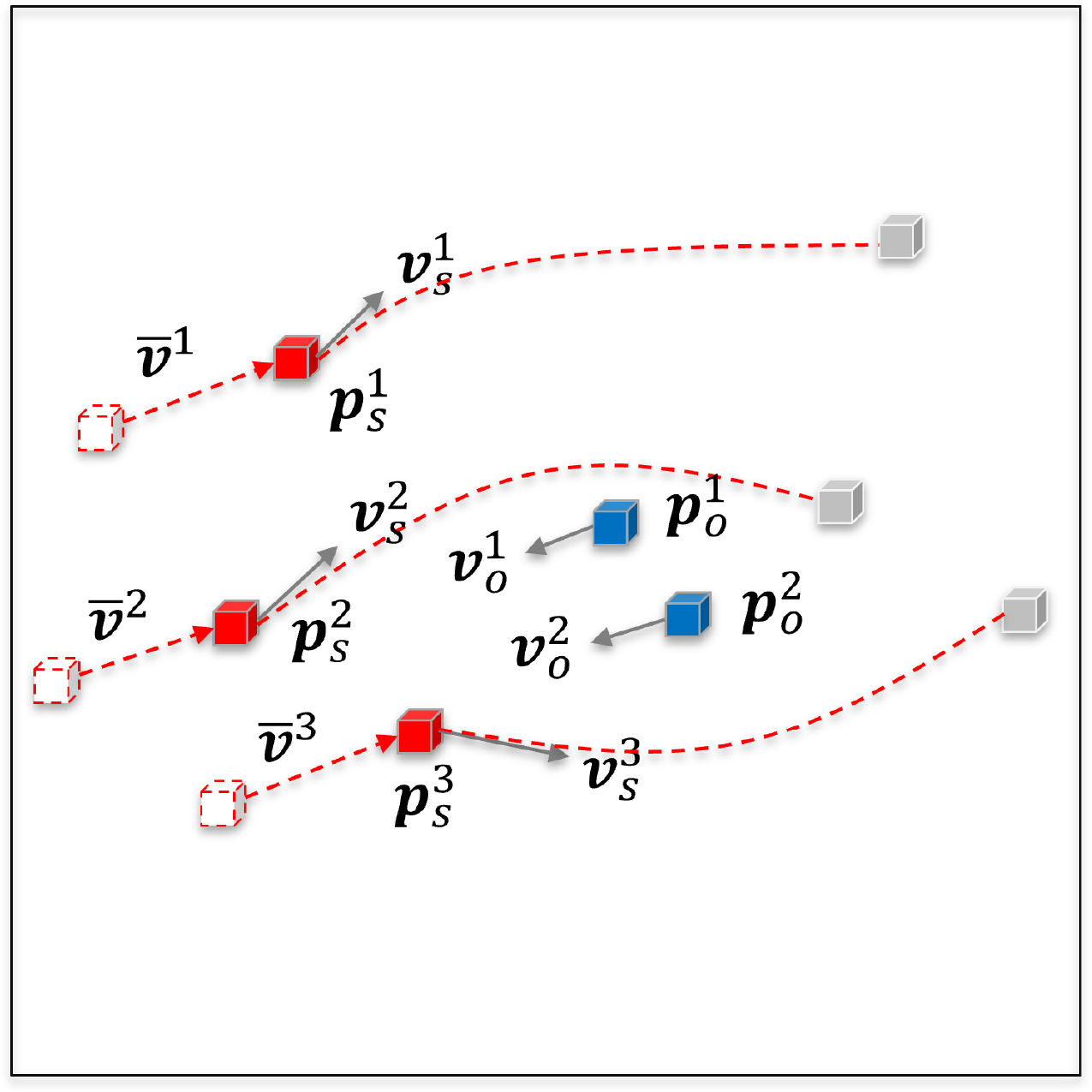} 
		\caption{} 
		\label{enva} 
	\end{subfigure} 
	\begin{subfigure}[b]{.48\linewidth}
		\centering
		\includegraphics[width=1.0\linewidth]{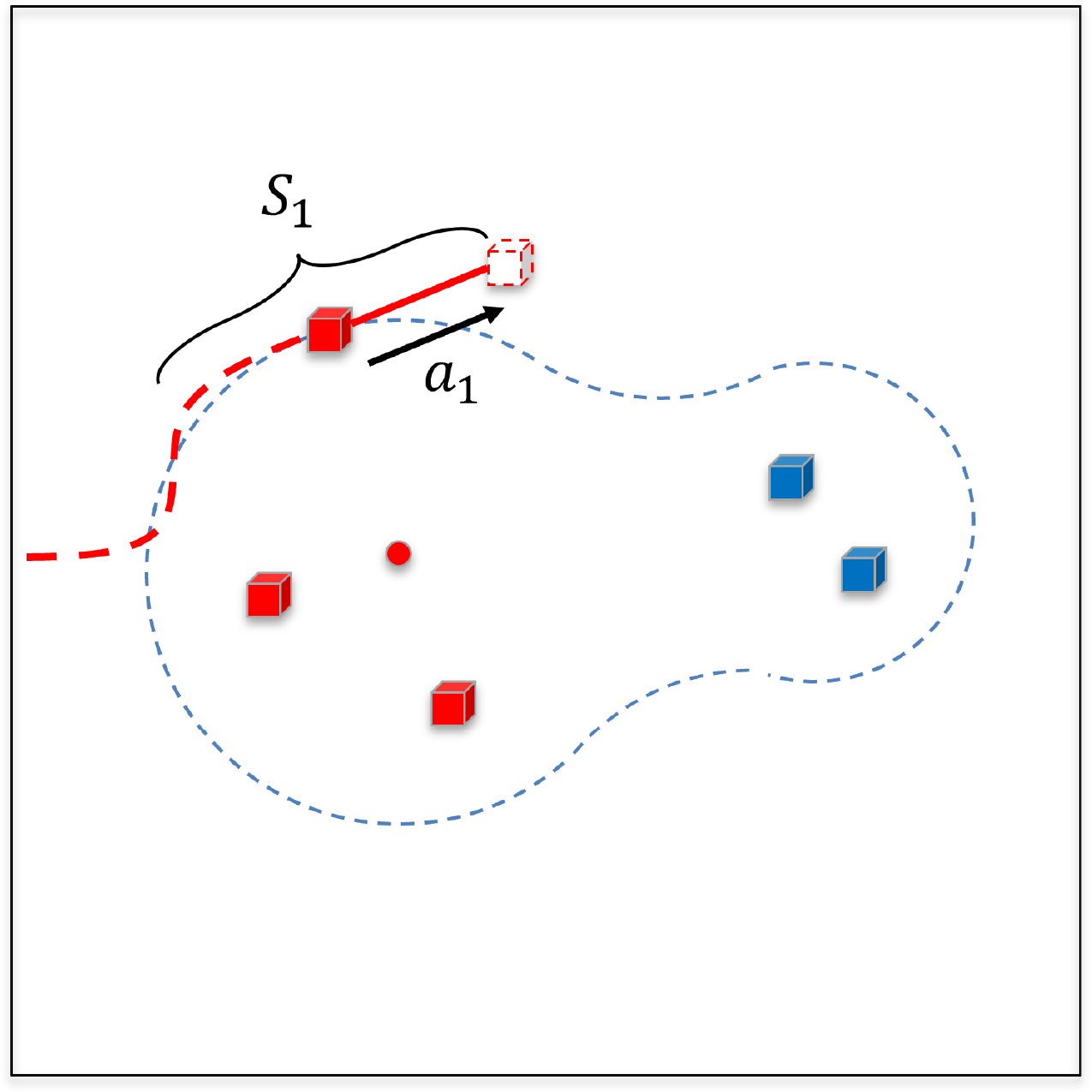} 
		\caption{} 
		\label{envb} 
	\end{subfigure} 
	\caption{Illustration of the environment. } 
	\label{env} 
\end{figure} 

%\begin{figure}[t] 
%	\centering 
%	\includegraphics[width=.6\linewidth]{figures/env2d.pdf} 
%	\caption{Illustration of the environment. } 
%	\label{env} 
%\end{figure} 

\subsection{Minimum Requirements on Observation Design} 

As illustrated in Equations \ref{eq:obstaclefield}, \ref{eq:swarmfield}, and \ref{eq:environmentfield}, the environmental potential field is determined solely by the position of the swarm's virtual center and the obstacles. Similarly, the field intensity only depends on the velocities of the swarm and the obstacles. Therefore, to achieve the innovative reward design, a UAV needs to observe at least the following properties: 
%its real-time position and velocity, the virtual center of its swarm, and the motion dynamics of obstacles. 
\begin{itemize}
	\item Self-property: $\{p_{s|x}, p_{s|y}, v_{s|x}, v_{s|y}\}$, with $p_x$, $v_x$ being its real-time position and velocity, and $|x$, $|y$ represent their $x-$ and $y-$ components, respectively. 
	\item Swarm property: $\{p_x^{*}, p_y^{*}, \bar{v}_x, \bar{v}_y\}$, with $p^s$ and $\bar{v}$ being position of the swarm's virtual center and the UAV's pre-planned velocities before avoiding collisions, respectively. 
	\item Obstacle property: $\{p_{o|x}, p_{o|y}, v_{o|x}, v_{o|y}\}$, with $p_{o}$ and $v_{o}$ being the obstacle's real-time position and velocity, respectively. 
\end{itemize} 
These properties are sufficient to represent the environment as the environment state is represented by the potential field which is solely constructed from the swarm and obstacle properties. Therefore, all information required for cooperation is embedded in the observation, eliminating the need for observation sharing and credit assignment schemes in network structures. 

Here, we present one possible observation design that meets the minimum requirements. We define the observation as a $(2 + m)\times 4$ array, with $m$ being the number of obstacles detected. The observation of each UAV can be seen as an array with $2 + m$ entries, each entry contains 4 elements. The first two entries contain the UAV's self-property and the swarm property, followed by $m$ entries of obstacle property. 

\section{Experiments} 

\begin{figure*}[t] 
	\centering 
	\includegraphics[width=1.\linewidth]{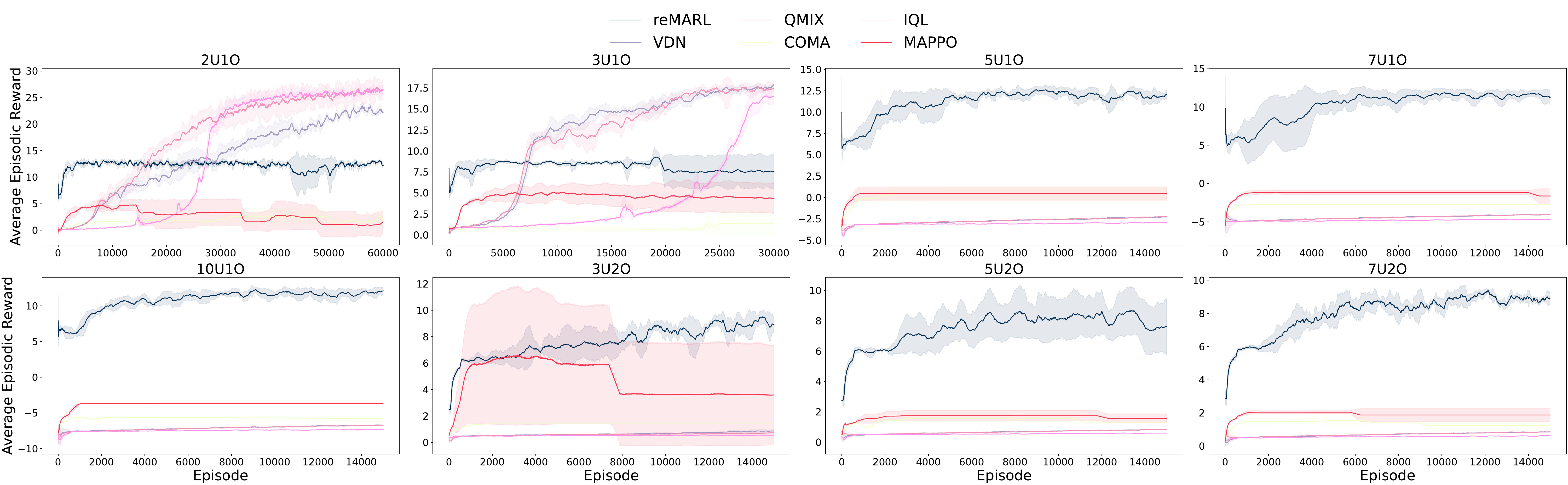} 
	\caption{Learning curves of \textit{reMARL} in comparison with state-of-the-art MARL algorithms. } 
	\label{result1} 
\end{figure*} 

\subsection{Environment} 

We developed a custom \textit{Gym}-like environment tailored to accurately simulate the cooperative collision avoidance of UAVs, as existing MARL environments are insufficient for closely modeling multi-UAV applications. The environment is depicted in Fig. \ref{enva}. Both the UAVs and obstacles are dynamic, moving within a square area of width $w$ and length $l$. In the illustration, the UAVs and obstacles are represented by red and blue cubes, respectively. The velocities of the UAVs and obstacles are indicated by gray arrows, while the target positions of the UAVs are marked as gray cubes. The pre-planned trajectories of the UAVs toward their respective target positions are illustrated with dashed curves. The environment operates episodically. At the start of each episode, the UAVs are spawned on the left side of the square, while the obstacles are placed on the right side. The UAVs initially follow their pre-planned trajectories and adjust their velocity directions to avoid collisions. The obstacles fly randomly toward the UAVs. The episode concludes under two conditions: either a collision occurs (with obstacle or between UAVs), or all UAVs successfully reach their respective target positions. Collisions occur when the distance between a UAV and an obstacle or between any two UAVs is smaller than a threshold $d_{col}$. 

Fig. \ref{envb} illustrates how the trajectories are generated from UAV's actions. In Fig. \ref{envb}, contours are plotted as blue dashed curves and the virtual center is marked by a red circle. The previous trajectories are represented by red dashed curves, while the newly generated trajectories from the immediate action are displayed as solid red curves. At each step, after UAV1 executes action $a_1$, the resulting trajectory is combined with the trajectories from the previous step to form $S_1$. This combined trajectory $S_1$ is then evaluated using Eq. \eqref{eq:fitness} to generate the reward signal. During an episode, the virtual center maintains its initial velocity and keeps moving accordingly. For smooth maneuvers, the UAVs action is defined as the change in velocity direction within the range $[-{\pi}/{4}, {\pi}/{4}]$. 

\subsection{Agent} 

Our method allows UAVs to learn cooperative behaviors solely by maximizing their individual reward signals, eliminating the need for complex credit assignment schemes or observation sharing mechanisms in agent's network. Therefore, the agent's learning is purely based on the DDPG algorithm, where each agent maintains a policy network and a value network. Techniques of experience replay and target networks \cite{double-q-learning} are also applied. 
%Due to the partial observability of UAVs, each agent's network can only take into its observations rather than the true state of the environment. 

%For continuous action space, the agent's learning is based on the DDPG algorithm \cite{ddpg}, where each agent maintains a policy network and a value network. Techniques of experience replay and target networks are also applied. Due to the partial observability of UAVs, each agent's network can only take into its observations rather than the true state of the environment. 

The policy network first processes the observation using a fully connected layer with 256 hidden units with a $ReLU$ activation function, and then an output layer with 1 unit with a $\tanh$ function. The last layer produces action of an agent $a=\pi_{\theta}(o)$. On the other hand, the value network first takes an agent's observations and action into two fully connected layers with 256 hidden units with $ReLU$ functions. Their outputs are concatenated and fed into another fully connected layer with 256 hidden units with a $ReLU$ function. Finally, the output is fed into an output layer with 1 unit, which generates the individual action value $Q(o, a)$. 
The parameters of the policy and value networks are optimized by minimizing the temporal difference loss in Eq. \ref{tdloss} and following the policy gradient in Eq. \ref{policygradient}, respectively. 

\begin{figure*}[htbp] 
	\centering
	\begin{subfigure}[b]{.35\linewidth}
		\centering
		\includegraphics[width=1.0\linewidth]{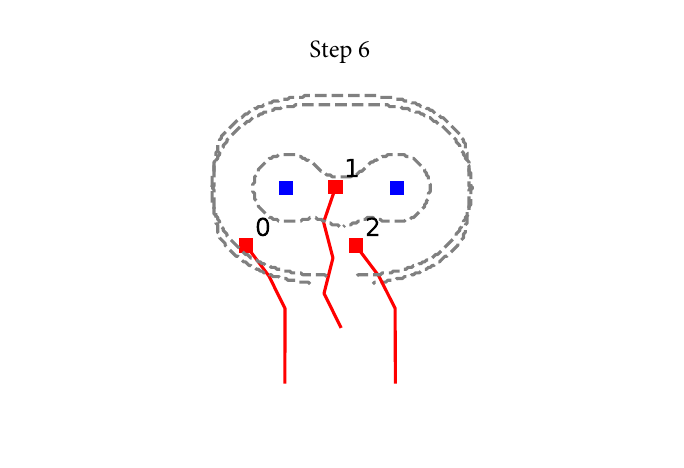} 
		\caption{} 
		\label{result2a} 
	\end{subfigure} 
	\begin{subfigure}[b]{.38\linewidth}
		\centering
		\includegraphics[width=1.0\linewidth]{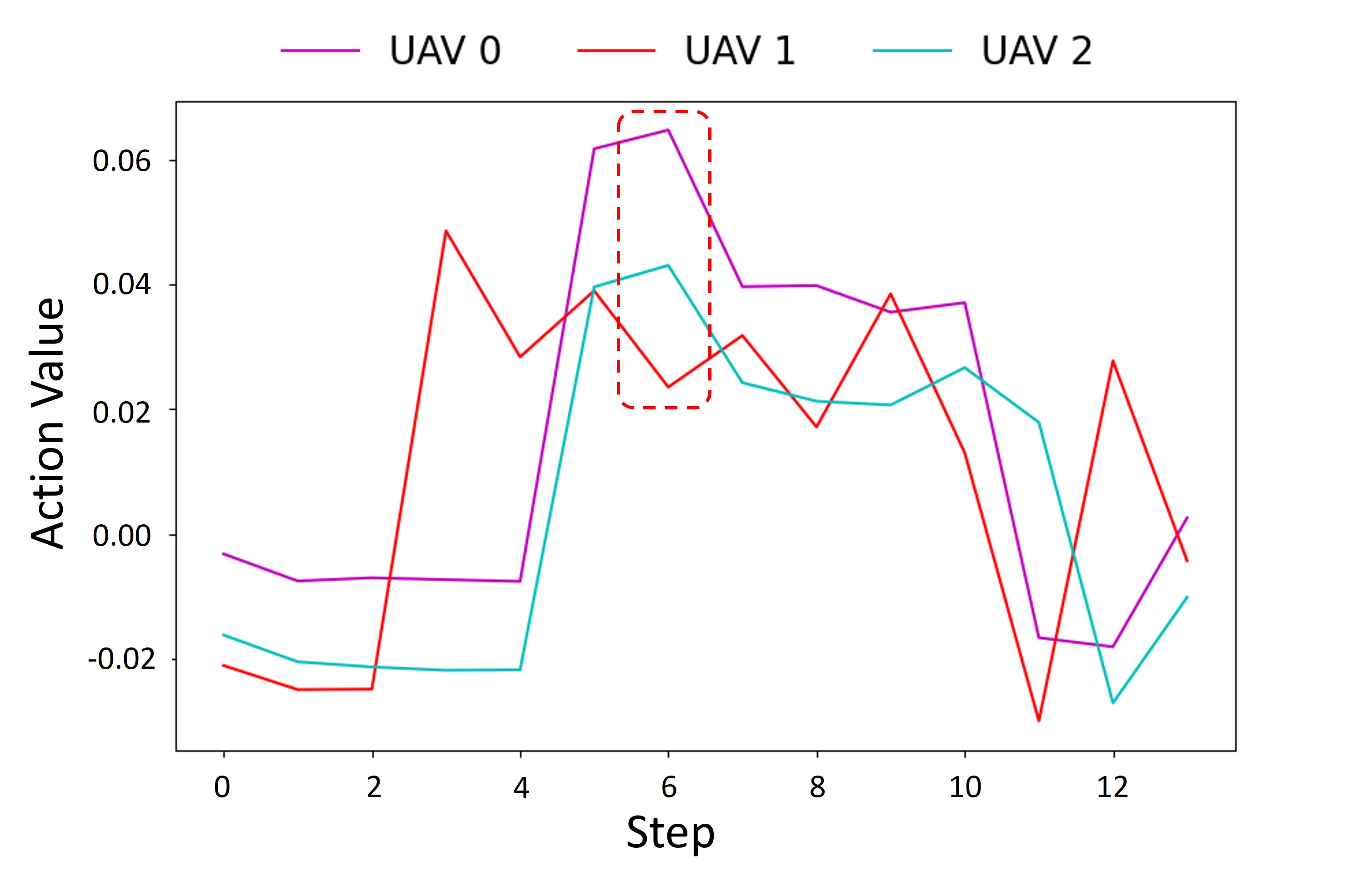} 
		\caption{} 
		\label{result2b} 
	\end{subfigure} 
	\caption{Trajectories and action values. } 
	\label{result2} 
\end{figure*} 

\subsection{Results} 
We compare with the following state-of-the-art MARL algorithms, such as COMA, VDN, QMIX, MAPPO, and IQL.  
%\begin{itemize}
%	\item \textbf{COMA}: COMA achieves explicit credit assignment based on the concept of difference rewards. 
%	\item \textbf{VDN}: VDN exploits implicit credit assignment schemes by decomposing the joint action value into local utility values for each agent additively. 
%	\item \textbf{QMIX}: QMIX adopts a similar approach with VDN. The difference is that the joint action value is decomposed into local utility values for each agent monotonically with non-negative weights. 
%	\item \textbf{IQL}: Agents learn independently while sharing policy and observations. 
%	\item \textbf{MAPPO}: All PPO agents maintain a centralized critic network to learn to cooperate. 
%\end{itemize} 
For the compared algorithms, the agents are trained only using the \textit{Swarming} part in Eq. \ref{eq:reward}. On the other hand, for our algorithm, the agents are trained using the complete form of Eq. \ref{eq:reward}, but its learning curve shows the \textit{Swarming} part only for comparison. 
For algorithms require a discrete action space, we sample 10 actions uniformly from $[-{\pi}/{4}, {\pi}/{4}]$ for each agent. 

The experiments are conducted in various scenarios: 2-UAV-1-Obstacle (\textit{2U1O}), 3-UAV-1-Obstacle (\textit{3U1O}), 5-UAV-1-Obstacle (\textit{5U1O}), 7-UAV-1-Obstacle (\textit{7U1O}), 10-UAV-1-Obstacle (\textit{10U1O}), 3-UAV-2-Obstacle (\textit{3U2O}), 5-UAV-2-Obstacle (\textit{5U2O}), and 7-UAV-2-Obstacle (\textit{7U2O}). The learning curves for all the algorithms are presented in Fig. \ref{result1}. As demonstrated, our approach \textit{reMARL} outperforms all baseline methods for swarm sizes greater than three and obstacle numbers larger than two. The superior performances are achieved through a combination of simple DDPG and the domain knowledge-driven reward. Since DDPG itself does not have any cooperation scheme, these performances are completely provided by the \textit{Contour} term in the reward signal. 
Notably, current state-of-the-art methods only exhibit superior performance in scenarios with very small swarm sizes (smaller than three). This suggests that UAVs need not follow contours when the swarm size is small, as safer and more energy-efficient trajectories exist. However, as the swarm size increases, adhering to contours becomes optimal to prevent collisions. 
%Moreover, COMA and MAPPO achieved the lowest average reward even in the simplest scenarios such as \textit{2U1O} and \textit{3U1O}, indicating the ineffectiveness of on-policy training. 

\begin{table*}[h]
	\centering
	\begin{tabular}{||c |c c c c||} 
		\specialrule{.1em}{.1em}{.1em}
		%		\multicolumn{6}{|c|}{\textit{Scenario 3U2O with }$r_{ave}$} \\
		\specialrule{.1em}{.1em}{.1em}
		& Reaction Time (Sec) & Energy Cost & \textit{min.} $d_{U2O}$ & \textit{min.} $d_{U2U}$ \\ 
		%		& Time (Sec) & Cost &  &  \\ [0.5ex] 
		\specialrule{.1em}{.1em}{.1em}
		\textit{Meta-Heuristic} & 0.48 $\pm$ 0.05  & 134.88 $\pm$ 8.3  & 38.70 & 40.31 \\ [0.5ex]  
		\textit{reMARL}  & \textbf{0.006 $\pm$ 0.03} & \textbf{19.72 $\pm$ 31.8 } & \textbf{24.78} & \textbf{29.45} \\ 
		\specialrule{.1em}{.1em}{.1em}
		{Improvement}  & 98.75\% & 85.37\% & 35.96\% & 26.94\% \\ [0.5ex] 
		\specialrule{.1em}{.1em}{.1em} 
		\specialrule{.1em}{.1em}{.1em}
	\end{tabular}
	\caption{Numerical results on safety, energy efficiency, and reaction time. }
	\label{table2}
\end{table*}

Fig. \ref{result1} shows that state-of-the-art credit assignment and observation sharing schemes are ineffective with large swarm sizes due to the curse of dimensionality and high computational complexity.
On the other hand, \textit{reMARL} works well even when the swarm size is as large as 10. This is because in \textit{reMARL}, the agents learn cooperative behaviors solely by maximizing their individual reward signals, eliminating the need for complex credit assignment schemes. Moreover, the observation sharing is only required when constructing environment field, decoupling it from network stuctures. The agents have global awareness through the swarm property in their observations. In this way, the interaction among agents is minimized enabling training for large swarm sizes.  

In addition, we also compare with meta-heuristics optimization approaches image processing domain knowledge \cite{huang2021}. As illustrated in Fig. \ref{result2}, UAVs gain the ability to adapt to complex environments where contours may be non-viable or non-existent using \textit{reMARL}. Fig. \ref{result2a} demonstrates the trajectories generated by \textit{reMARL}, where the UAVs and obstacles are represented by red and blue cubes, respectively. 
%The virtual leader is depicted with red dot. 
The gray dashed curves depict the contours on the environment field, which are used as trajectories in pure meta-heuristic optimization algorithms. In this case, UAV 1 and 2 in the middle would have to make a detour to circumvent the obstacles if strictly follow meta-heuristic optimization algorithms and lead to lengthy trajectories and sharp turnings. Therefore, the contours are not viable trajectories as they did not maximize energy efficiency of UAVs while ensuring safety. However, using \textit{reMARL}, the UAVs choose to navigate between the obstacles, resulting in a shorter, smoother and more energy-efficient trajectory while maintaining safety, instead of deviating. This is because traditional meta-heuristics only utilize the \textit{Contour} part as cost function, which essentially follows the active contour model in contour extraction. In contrast, \textit{reMARL} employs an integration of both \textit{Contour} and \textit{Swarming} parts, where the \textit{Swarming} part takes care of safety, formation and efficiency. These two parts work together to generate more efficient trajectories. Fig. \ref{result2b} shows the action value $Q$ at each step. UAV 0 has the highest action value at step 6 as it strictly follows the contours which guarantee no collision. UAV 1 and 2, on the other hand, show slightly lower action values than UAV 0, but still larger than other steps of their trajectories. This shows that UAV 1 and 2 know for sure that passing in between the two obstacles is safe, although deviating from the contours. 

%Moreover, the reaction time is substantially reduced, as actions are determined through a simple forward pass of neural networks, rather than heuristic-based search. 
Numerical results are presented in Table \ref{table2}, assuming $d_{col}=20$. We use the average curvature of trajectories to evaluate the UAV's energy costs, which is defined in Eq. \ref{curvature}.  
\begin{equation} \label{curvature} 
	\begin{aligned} 
		E_n = \int \mid S''(p) \mid dp, 
	\end{aligned} 
\end{equation} 
where $p$ is arc length parameter, $S(p)=[x(p), y(p)]$ denotes a waypoint on the trajectory, and $S''(p)$ denotes the second order derivative of $S(p)$. 

Table \ref{table2} shows that \textit{reMARL} achieves 98.75\% shorter reaction time and 85.37\% lower energy costs compared to conventional contour-based meta-heuristic search. This is achieved by smoother trajectories that are 35.97\% closer to obstacles and 25.94\% closer to their teammates. Although the distances are smaller, safety is still guaranteed as they are larger than the threshold $d_{col}=20$. This is achieved by the \textit{Swarming} part of the reward and the PSO search for contour adjustment. The result in Table \ref{table2} are acquired on Intel i7 processor with a base frequency of 2.50 $GHz$ and 32 $GB$ memory.

%\begin{enumerate}
%	\item UAVs gain the ability to adapt to complex environments where contours may be non-viable or non-existent as shown in Fig. \ref{result2}; 
%	\item The reaction time is substantially reduced, as actions are determined through a simple forward pass of neural networks. 
%\end{enumerate}

%\begin{figure*}[htbp] 
%	\centering
%	\begin{subfigure}[b]{.48\linewidth}
	%		\centering
	%		\includegraphics[width=1.0\linewidth]{figures/learning_curve_2u1o.pdf} 
	%		\caption{2-UAV swarm avoiding 1 obstacle. } 
	%		\label{result1a} 
	%	\end{subfigure} 
%	\begin{subfigure}[b]{.48\linewidth}
	%		\centering
	%		\includegraphics[width=1.0\linewidth]{figures/learning_curve_3u1o.pdf} 
	%		\caption{3-UAV swarm avoiding 1 obstacle. } 
	%		\label{result1b} 
	%	\end{subfigure} 
%	\begin{subfigure}[b]{.48\linewidth}
	%		\centering
	%		\includegraphics[width=1.0\linewidth]{figures/learning_curve_5u1o.pdf} 
	%		\caption{5-UAV swarm avoiding 1 obstacle. } 
	%		\label{result1c} 
	%	\end{subfigure} 
%	\begin{subfigure}[b]{.48\linewidth}
	%		\centering
	%		\includegraphics[width=1.0\linewidth]{figures/learning_curve_10u1o.pdf} 
	%		\caption{10-UAV swarm avoiding 1 obstacle. } 
	%		\label{result1d} 
	%	\end{subfigure} 
%	\caption{Learning curves of our method in comparison with state-of-the-art MARL algorithms. } 
%	\label{result1} 
%\end{figure*} 

\section{Conclusion} 

We propose \textit{reMARL} - an innovative MARL algorithm leveraging domain knowledge-driven reward for energy-efficient and cooperative collision avoidance of UAV swarms. Extensive experiments are conducted to proof that \textit{reMARL} achieves stabler and higher performance, compared to the state-of-the-art MARL algorithms, even with large swarm sizes. At the same time, \textit{reMARL} achieves 98.75\% shorter reaction time and 85.37\% lower energy costs compared to conventional meta-heuristic search based on the same domain knowledge, while ensuring safety. 

\bibliography{aaai25}

\end{document}